\documentclass[preprint,12pt]{elsarticle}

\usepackage{amssymb}

\usepackage{lineno}
\RequirePackage{xspace}
\usepackage{relsize}
\def\babar{\mbox{\slshape B\kern-0.1em{\smaller A}\kern-0.1em
    B\kern-0.1em{\smaller A\kern-0.2em R}}}
\def\CP     {\ensuremath{C\!P}\xspace}
\def\en     {\ensuremath{e^-}\xspace}
\def\ep     {\ensuremath{e^+}\xspace}
\def\Bbar   {\kern 0.18em\overline{\kern -0.18em B}{}\xspace}
\def\Dbar   {\kern 0.2em\overline{\kern -0.2em D}{}\xspace}
\def\Kbar   {\kern 0.2em\overline{\kern -0.2em K}{}\xspace}
\def\jpsi   {\ensuremath{{J\mskip -3mu/\mskip -2mu\psi\mskip 2mu}}\xspace}
\def\psitwos{\ensuremath{\psi{(2S)}}\xspace}
\def\KS     {\ensuremath{K^0_{\scriptscriptstyle S}}\xspace} 
\mathchardef\Upsilon="7107
\def\Y#1S{\ensuremath{\Upsilon{(#1S)}}\xspace}
\def\YnS{\ensuremath{\Upsilon{(nS)}}\xspace}

\newcommand {\mev}{\ensuremath{{\mathrm{\,Me\kern -0.1em V}}}\xspace}
\newcommand {\mevcc}{\ensuremath{{\mathrm{\,Me\kern -0.1em V\!/}c^2}}\xspace}
\newcommand {\gevtwocc}{\ensuremath{{\mathrm{\,Ge\kern -0.1em V^{2}\!/}c^4}}\xspace}
\def\invab  {\ensuremath{\mbox{\,ab}^{-1}}\xspace}
\def\invfb  {\ensuremath{\mbox{\,fb}^{-1}}\xspace}
\newcommand{\stat}{\ensuremath{\mathrm{(stat)}}\xspace}
\newcommand{\syst}{\ensuremath{\mathrm{(syst)}}\xspace}

\journal{Nuclear Physics B}

\begin{document}

\begin{frontmatter}


\title{Recent results on hot topics from Belle}


\author{Gagan B. Mohanty}

\address{Tata Institute of Fundamental Research, Homi Bhabha Road,\\
         Colaba, Mumbai 400 005, INDIA}

\begin{abstract}
We report a sample of recent and topical physics results based on the
data recorded with the Belle detector at the KEK $B$-factory in Japan.
\end{abstract}


\end{frontmatter}


\section{Introduction}
\label{Sec-1}
The Belle experiment is a multitasking magnetic spectrometer that
operated for more than a decade at the KEKB asymmetric-energy
$\ep\en$ collider. Before closing down in June 2010 to make way
for its upgrade (Belle II), it has succeeded in collecting a
world-record sample of data over $1\invab$ near various $\YnS$
resonances. We present herein a sample of recent interesting
results from Belle based on its full statistics.

\section{\boldmath Charm mixing and $\CP$ violation}
\label{Sec-2}
Under weak interaction a flavored neutral meson can be represented as
a two-state quantum system, leading to possible transition between
the two states. The phenomenon going by the name {\em neutral meson
mixing} is intimately related to the difference between the flavor
and mass eigenstates of the meson-antimeson system. Such mixing is
already well established for $K^0$, $B^0$ and $B^0_s$ mesons. The
observed mixing rates are consistent with the standard model (SM)
predictions that depend on the Cabibbo-Kobayashi-Maskawa (CKM)
matrix~\cite{ckm} elements, appearing in the short-distance box
diagrams for mixing. For $D^0$ mesons the box diagrams are, however,
both Cabibbo and GIM~\cite{gim} suppressed giving rise to a small
contribution. Consequently, $D^0$-$\Dbar^0$ mixing is dominated by
long-distance processes that are difficult to calculate. Theoretical
estimates for the mixing parameters $x\equiv\Delta m/\Gamma$ and
$y\equiv\Delta\Gamma/2\Gamma$ range over two to three orders of
magnitude~\cite{mix}. Here, $\Delta m$ and $\Delta\Gamma$ are the mass
and decay width differences between the two $D$ mass eigenstates, and
$\Gamma$ is their average decay width.

Violation of charge-parity ($\CP$) symmetry in the charm sector provides
an interesting test for new physics (NP) as the SM predicts a very small
asymmetry owing to: a) large Cabibbo and GIM suppression, similar to
mixing, and b) a lack of large hierarchy in the down-type quark masses.
Among various $D$ decay modes the singly Cabibbo-suppressed (SCS) decays
constitute the most promising candidate to probe $\CP$ violation~\cite{scs}
with typical SM values are of the order of $10^{-3}$. With respect to
such percentage effects one needs a good control on the theory
prediction, something that is in general lacking in the charm sector due
to long-distance effects. Furthermore, with $D^0$-$\Dbar^0$ mixing being
firmly established (see below) one would like to explore possible $\CP$
violation in mixing or due to interference between mixing and decay
amplitudes.

\section{\boldmath Mixing in $D\to K\pi$ decays}
\label{Sec-3}
We search for $D^0$-$\Dbar^0$ mixing in $D\to K\pi$ decays~\cite{res1}
by taking a ratio of the time-dependent rate of the wrong-sign (WS) decay
$D^0\to K^+\pi^-$ to that of the right-sign (RS) decay $D^0\to K^-\pi^+$.
The RS and WS processes are identified via the decay chain $D^{\star +}
\to D^0(K^{\mp}\pi^{\pm})\pi^+_s$, where one compares the charge of the
slow pion $\pi_s$ with that of the pion arising from the $D$ decay. If
the two are the same, it would be a RS decay; else, a WS decay. The RS
amplitude is dominated by a Cabibbo-favored (CF) decay $D^0\to K^-\pi^+$
with a negligible contribution from $D^0$-$\Dbar^0$ mixing followed by a
doubly Cabibbo-suppressed (DCS) decay $\Dbar^0\to K^-\pi^+$. In contrast,
for the WS case the two contributing amplitudes, from the DCS decay $D^0
\to K^+\pi^-$ and $D^0$-$\Dbar^0$ mixing followed by the CF decay
$\Dbar^0\to K^+\pi^-$, are of similar magnitude. The time-dependent
ratio of WS to RS decay rates is thus given by
\begin{eqnarray}
R(t/\tau)\equiv\frac{\Gamma_{\rm WS}(t/\tau)}
{\Gamma_{\rm RS}(t/\tau)}\approx R_D+\sqrt{R_D}y^{\prime}
\frac{t}{\tau}+\frac{x^{\prime 2}+y^{\prime 2}}{4}
\left(\frac{t}{\tau}\right)^2
\label{eq1}
\end{eqnarray}
to the second order in mixing parameters. Here, $t$ is the proper decay
time, $\tau$ is the average $D^0$ lifetime, $R_D$ is the ratio of DCS
to CF decay rates, $x^{\prime}=x\cos\delta+y\sin\delta$, and $y^{\prime}
=-x\sin\delta+y\cos\delta$ are the ``rotated'' mixing parameters with
$\delta$ being the strong phase difference between the DCS and CF decay
amplitudes.

\begin{table}[!ht]
\caption{Results of the time-dependent fit to $R(t/\tau)$, where
DOF denotes the number of degrees of freedom. The uncertainties
are statistical and systematic combined.}
\begin{center}
\begin{tabular}{lccc}
\hline\hline
Test hypothesis & $\chi^2$/DOF & Parameter & Fit result ($10^{-3}$)\\
\hline
Mixing          & $4.2/7$ & $R_D$      & $3.53\pm0.13$\\
                &         & $y^\prime$ & $4.6\pm3.4$\\
                &         & $x^{\prime 2}$ & $0.09\pm0.22$\\
\hline
No mixing      & $33.5/9$ & $R_D$       & $3.864\pm0.059$\\
\hline\hline
\label{tab2}
\end{tabular}
\end{center}
\end{table}
Based on a $976\invfb$ of data sample, we select $2,980,710\pm1885$ RS and
$11,478\pm177$ WS decay candidates. The proper decay time of these decay
candidates is calculated as $t=m_{D^0}\vec{L}\cdot\vec{p}/|\vec{p}|^2$,
where $\vec{L}$ is a vector joining the $D^0$ production and decay
vertices, $\vec{p}$ is the $D^0$ momentum, and $m_{D^0}$ is the nominal
$D^0$ mass~\cite{pdg}. Due care has been taken in incorporating detector
resolution effects to the measured $t$ distribution. We fit the obtained
time-dependent decay rate ratios according to the expression in
Eq.~(\ref{eq1}). Two hypotheses, with and without $D^0$-$\Dbar^0$ mixing,
are tested and the corresponding results are given in Table~\ref{tab2}.
The $\chi^2$ difference between the two is $29.3$ for $2$ degrees of
freedom, corresponding to a probability of $4.3\times10^{-7}$. This
means, the no-mixing hypothesis is excluded at a level of $5.1$ standard
deviations ($\sigma$). Our results constitute the first observation of
$D^0$-$\Dbar^0$ mixing from a single $\ep\en$ collider experiment, and
are in agreement with those from experiments at the hadron
machine~\cite{hadron}.

\section{\boldmath Mixing and $\CP$ violation in $D\to\KS\pi^+\pi^-$}
\label{Sec-3}
We simultaneously probe mixing and $\CP$ violation~\cite{res2} by
studying time dependence of the Dalitz plot in the self-conjugated decay
$D^0\to\KS\pi^+\pi^-$ (denoted by $f$ below). For a given point in the
Dalitz plot [$m^2_+\equiv m^2(\KS\pi^+)$, $m^2_-\equiv m^2(\KS\pi^-)$],
the decay amplitude is
\begin{eqnarray}
{\cal M}(t)={\cal A}_f(m^2_+,m^2_-)\frac{{\rm e}_1(t)+
{\rm e}_2(t)}{2}+\frac{q}{p}{\cal A}_{\bar{f}}(m^2_+,m^2_-)
\frac{{\rm e}_1(t)-{\rm e}_2(t)}{2},
\label{eq2}
\end{eqnarray}
where $q,p$ are the complex coefficients that relate mass to flavor
eigenstates, and ${\rm e}_{1,2}(t)={\rm e}^{-(im_{1,2}+\Gamma_{1,2}/2)t}$
with $m_{1,2}$ and $\Gamma_{1,2}$ being the mass and decay width of the
mass eigenstates. The first term is the time-dependent amplitude for
the (direct) decay $D^0\to\KS\pi^+\pi^-$, and the second one is the
amplitude for $D^0$-$\Dbar^0$ mixing followed by $\Dbar^0\to\KS\pi^+\pi^-$.
Taking the modulus squared of Eq.~(\ref{eq2}) gives the time-dependent
$D^0$ decay rate as
\begin{eqnarray}
|{\cal M}(t)|^2&=&\frac{{\rm e}^{-\Gamma t}}{2}\{(|{\cal A}_f|^2+
\left|\frac{q}{p}\right|^2|{\cal A}_{\bar{f}}|^2)\cosh(\Gamma yt)
+(|{\cal A}_f|^2-\left|\frac{q}{p}\right|^2|{\cal A}_{\bar{f}}|^2)\\\nonumber
&&\cosh(\Gamma xt)+2\,{\cal R}(\frac{q}{p}{\cal A}_{\bar{f}}A^{\star}_f)
\sinh(\Gamma yt)-2\,{\cal I}(\frac{q}{p}{\cal A}_{\bar{f}}A^{\star}_f)
\sinh(\gamma xt)\}.
\label{eq3}
\end{eqnarray}
The equivalent quantity in the $\Dbar^0$ case also contains terms
proportional to $\cosh(\Gamma xt)$, $\cosh(\Gamma yt)$, $\sinh(\Gamma xt)$
and $\sinh(\Gamma yt)$. Thus, by fitting the time-dependent $D^0$ and
$\Dbar^0$ decay rates we can determine mixing parameters $x$ and $y$ as
well as look for mixing-induced $\CP$ violation by checking whether $|q/p|$
deviates from unity or Arg($q/p$) from zero, and if so, by how much.

We use an {\em isobar model}~\cite{isobar} to describe the decay
amplitude ${\cal A}_f(m^2_+,m^2_-)$ as $\sum_ja_j{\rm e}^{i\delta_j}A_j$,
where $a_j$ and $\delta_j$ are the magnitude and phase of a given
intermediate state $j$, and $A_j$ is the product of a relativistic
Breit-Wigner function and Blatt-Weisskopf form factors. Thus in the
Dalitz-plot fit we obtain amplitudes and phases of various intermediate
states, that lead to the final state of $\KS\pi^+\pi^-$, in addition
to mixing parameters $x$ and $y$. For no direct $\CP$ violation,
${\cal A}_f(m^2_+,m^2_-)={\cal A}_{\bar{f}}(m^2_-,m^2_+)$. Otherwise,
one needs to separately consider ($a_j,\delta_j$) for $D^0$ and
($\bar{a}_j,\bar{\delta}_j$) for $\Dbar^0$ decays.

Using a data sample of $921\invfb$ recorded near the $\YnS$
($n=4,5$) resonances with Belle, we select $1,231,731\pm
1633$ signal candidates with a purity above $95\%$. We perform a
time-integrated fit to the Dalitz plot distributions of these events
by varying the amplitudes and phases for different intermediate
states, separately for $D^0$ and $\Dbar^0$ decays. As the two sets
of parameters are found to be consistent within uncertainties, we
set ${\cal A}_f(m^2_+,m^2_-)={\cal A}_{\bar{f}}(m^2_-,m^2_+)$. Of course,
this assumption of no direct $\CP$ violation has no bearing for $\CP$
violation related to mixing. Next we follow a two-step fit procedure.
In the first step, we fit to a combined sample of $D^0$ and $\Dbar^0$
with the fit observables ($x,y$), the $D^0$ lifetime $\tau$, some
timing resolution parameters, and isobar model parameters ($a_j,
\delta_j$). In the second step, $\CP$ violation related to mixing
is allowed resulting in two more fit observables: $|q/p|$ and $\phi=$
Arg($q/p$). Results of the above two fits are listed in Table~\ref{tab3}.
The two sets of mixing parameters obtained are pretty much identical,
and they constitute a $2.5\sigma$ evidence for $D^0$-$\Dbar^0$ mixing.
With $|q/p|$ and $\phi$ being consistent with unity and zero, respectively,
there is no $\CP$ violation either in $D^0$-$\Dbar^0$ mixing or due to
interference between mixing and decay.
\begin{table}[!ht]
\caption{Results for mixing parameters $x$ and $y$ from the no and
with $\CP$ violation (CPV) case. The uncertainties are respectively
statistical, experimental systematic, and the error due to amplitude
model.}
\begin{center}
\begin{tabular}{lcc}
\hline\hline
Fit type & Parameter & Fit result \\
\hline
No CPV   & $x\,(\%)$ & $0.56\pm0.19^{+0.03}_{-0.09}\,^{+0.06}_{-0.09}$\\
         & $y\,(\%)$ & $0.30\pm0.15^{+0.04}_{-0.05}\,^{+0.03}_{-0.06}$\\
With CPV & $x\,(\%)$ & $0.56\pm0.19^{+0.04}_{-0.08}\,^{+0.06}_{-0.08}$\\
         & $y\,(\%)$ & $0.30\pm0.15^{+0.04}_{-0.05}\,^{+0.03}_{-0.06}$\\
         & $|q/p|$ & $0.90^{+0.16}_{-0.15}\,^{+0.05}_{-0.04}\,^{+0.06}_{-0.05}$\\
         &$\phi\,(^\circ)$ & $-6\pm11\pm3\,^{+3}_{-4}$\\
\hline\hline
\label{tab3}
\end{tabular}
\end{center}
\end{table}

\section{\boldmath Search for $\CP$ violation in $D^0\to\pi^0\pi^0$}
\label{Sec-4}

The $D^0\to\pi^0\pi^0$ decay, a typical SCS one, is expected to
exhibit large $\CP$ violation in several NP models such as those
with large chromomagnetic dipole operators~\cite{cheng} and the
triplet model~\cite{hiller}. The enhanced $\CP$ asymmetry value
could lie anywhere between $1\%$ and $8\%$. Further motivation
for our study came from an early measurement by LHCb~\cite{lhcb},
supported by CDF~\cite{cdf}, suggesting a $3.5\sigma$ effect on the
difference of direct $\CP$ asymmetries ($\Delta A_{\CP}$) between
$D^0\to K^+K^-$ and $D^0\to\pi^+\pi^-$ decays. Though the current
world average for $\Delta A_{\CP}$ stands $2.3\sigma$ away from
zero~\cite{hfag}, there exists a good deal of theoretical
motivation~\cite{cheng,hiller,bhuban} to look for $\CP$ violation
in $D^0\to\pi^0\pi^0$. Experimentally, the previous result from
CLEO~\cite{cleo} was consistent with zero with an uncertainty of
$4.8\%$ -- as big as the NP prediction.

We measure the time-integrated $\CP$ asymmetry in the $D^0\to\pi^0\pi^0$
and $D^0\to\KS\pi^0$ decays~\cite{res3} using $966\invfb$ of data collected
with Belle. In the process $D^{\star +}\to D^0\pi^+_s$ coming from $\ep\en
\to c\bar{c}$, the charge of the low-momentum or ``slow'' pion $\pi^+_s$
determines the flavor of the neutral charm meson (whether it is a $D^0$
or a $\Dbar^0$). The measured asymmetry
\begin{eqnarray}
A_{\rm rec}=\frac{N^{D^{\star +}\to D^0\pi^+_s}_{\rm rec}-
N^{D^{\star -}\to\Dbar^0\pi^-_s}_{\rm rec}}{N^{D^{\star +}\to
D^0\pi^+_s}_{\rm rec}+N^{D^{\star -}\to\Dbar^0\pi^-_s}_{\rm rec}},
\label{eq4}
\end{eqnarray}
where $N^i_{\rm rec}$ is the number of reconstructed events tagged as
$i$, has three contributions: the underlying $\CP$ asymmetry $A_{\CP}$,
the forward-backward asymmetry ($A_{\rm FB}$) due to $\gamma$-$Z^\star$
interference in $\ep\en\to c\bar{c}$ and higher-order QED processes,
and the detection asymmetry between positively and negatively charged
pions ($A^{\pi_s}_{\epsilon}$). We subtract the $A_{\rm rec}$ measured
in the CF decay $D^0\to K^-\pi^+$ (``untagged'') from $D^{\star +}\to D^0
\pi^+_s; D^0\to K^-\pi^+$ (``tagged'') to estimate $A^{\pi_s}_{\epsilon}$.
The implicit assumption here is that both $D^\star$ and $D$ mesons have
the same $A_{\rm FB}$ value. After correcting for $A^{\pi_s}_{\epsilon}$,
one is left with
\begin{eqnarray}
A^{\rm cor}_{\rm rec}=A_{\CP}+A_{\rm FB}(\cos\theta^\star),
\label{eq4}
\end{eqnarray}
where $\theta^\star$ is the $D^{\star +}$ polar angle in the center
of mass (CM) frame. While $A_{\CP}$ is independent of kinematics,
$A_{\rm FB}$ is an odd function of $\cos\theta^\star$. Making use of
this important distinction, we obtain $A_{\CP}=[A^{\rm cor}_{\rm rec}
(\cos\theta^\star)+A^{\rm cor}_{\rm rec}(-\cos\theta^\star)]/2$ and
$A_{\rm FB}=[A^{\rm cor}_{\rm rec}(\cos\theta^\star)-A^{\rm cor}_{\rm rec}
(-\cos\theta^\star)]/2$.

Based on a total signal yield of $34,460\pm273$ $D^0\to\pi^0\pi^0$ events, we
obtain $A_{\CP}=[-0.03\pm0.64\stat\pm0.10\syst]\%$, which improves over the
previous results~\cite{cleo} by an order of magnitude and shows no hint for
$\CP$ violation. We also measure $A_{\CP}(D^0\to\KS\pi^0)=[-0.21\pm0.16\stat
\pm0.07\syst]\%$, which supersedes Belle's earlier result~\cite{byongrok}.
After subtracting $\CP$ violation due to $K^0$-$\Kbar^0$ mixing, the $\CP$
asymmetry for $D^0\to\Kbar^0\pi^0$ is found to be $(+0.12\pm0.16\pm0.07)\%$,
again consistent with no $\CP$ violation.

\section{\boldmath Lepton forward-backward asymmetry in $B\to X_S\ell^+\ell^-$}
\label{Sec-5}

The flavor-changing neutral current (FCNC) transition $b\to s\ell^+\ell^-$ ($\ell=
e,\mu$) is forbidden at tree level in the SM. However, it can occur at higher order
via electroweak loop (penguin) and $W^+W^-$ box diagrams. The corresponding decay
amplitudes are expressed in terms of the effective Wilson coefficients~\cite{wilson}:
$C_7$ for the electromagnetic penguin, $C_9$ and $C_{10}$ for the vector and
axial-vector electroweak contributions, respectively~\cite{buchalla}. In presence
of NP contributions, these coefficients are expected to differ from SM predictions,
leading to a dramatic change in the decay rate and angular distributions of the $b
\to s\ell^+\ell^-$ transition~\cite{np-sll}. For instance, the lepton forward-backward
asymmetry in the $B\to X_S\ell^+\ell^-$ decays,
\begin{eqnarray}
A_{\rm FB}=\frac{\Gamma(B\to X_S\ell^+\ell^-;\cos\theta>0)-\Gamma(B\to X_S\ell^+\ell^-;\cos\theta<0)}
{\Gamma(B\to X_S\ell^+\ell^-;\cos\theta>0)+\Gamma(B\to X_S\ell^+\ell^-;\cos\theta<0)},
\label{eq5}
\end{eqnarray}
where $\theta$ is the angle between the $\ell^+ (\ell^-)$ and the $B$ meson momentum
in the $\ell^+\ell^-$ CM frame and $X_S$ is a hadronic system containing an $s$ quark,
exhibits an excellent sensitivity for physics beyond the SM.

Motivated by this, we perform the first measurement of $A_{\rm FB}$ in the inclusive
$B\to X_S\ell^+\ell^-$ decays as a function of the dilepton invariant mass squared
$q^2=m^2_{\ell^+\ell^-}$~\cite{res4}, using $772\times10^6$ $B\Bbar$ pairs collected
at the $\Y4S$ resonance. By inclusive here one means a sum of several exclusive
hadronic final states representing the $X_S$ system. We reconstruct $B$ mesons in
$18$ hadronic final states with $X_S\equiv\{K\}\{n\pi\}$, $K=K^{\pm},\KS$ and $n=1...4$
of which at most one pion can be neutral, together with two oppositely charged
leptons (electrons or muons). In case of $B^0$ ($\Bbar^0$) decays, only self-tagging
modes with a $K^+$ ($K^-$) are utilized. Signal events are identified with two
kinematic variables, calculated in the $\Y4S$ rest frame: the beam-energy constrained
mass $M_{\rm bc}=\sqrt{E^{2}_{\rm beam}-|\vec{p}_B|^2}$ and the energy difference $\Delta
E=E_B-E_{\rm beam}$, where $E_{\rm beam}$ is the beam energy, and ($E_B,\vec{p}_B$)
are the reconstructed energy and momentum of the $B$-meson candidate. To reduce
contamination from the $\ep\en\to q\bar{q}$ ($q=u,d,s,c$) continuum background, we use
a neural network mostly based event topology and $B$ vertex-fit quality. Charmonia
contributions from $B\to X_S\jpsi[\psitwos]$ with $\jpsi[\psitwos]\to\ell^+\ell^-$, are
suppressed by rejecting (``vetoing'') events with a dilepton invariant mass in the
following two ranges: $-400$ to $150\mevcc$ ($-250$ to $100\mevcc$) and $-250$ to
$100\mevcc$ ($-150$ to $100\mevcc$)  around the nominal $\jpsi$ and $\psitwos$
mass~\cite{pdg} mass for the electron (muon) channel.

In total, $140\pm19\stat$ $B\to X_S\en\ep$ and $161\pm20\stat$ $B\to X_S\mu^+\mu^-$
signal candidates are selected in the data sample. To study the $q^2$ dependence of
$A_{\rm FB}$, we divide the data into four $q^2$ bins: $[0.2,4.3]$, $[4.3,7.3(8.1)]$,
$[10.5(10.2),11.8(12.5)]$, and $[14.3,25.0]\gevtwocc$ for the electron (muon) channel,
where the gap regions correspond to the veto described earlier. The $A_{\rm FB}$ value
is found to be consistent with the SM prediction in the two intermediate $q^2$ bins,
while it deviates from the SM in the lowest $q^2$ bin by $1.8\sigma$ (including the
systematic uncertainty). Results in the last two bins exclude $A_{\rm FB}<0$ at a
$2.3\sigma$ level.

\section{\boldmath Observation of the decay $B^0\to\eta^{\prime}K^{\star}(892)^0$}
\label{Sec-6}

The $B^0\to\eta^{\prime}K^{\star}(892)^0$ decay proceeds via the $b\to s$ penguin
and Cabibbo-suppressed $b\to u$ tree diagrams. There is a destructive interference
between two contributing amplitudes leading to a small decay branching fraction.
The interference pattern could also give rise to a large direct $\CP$ violation.
Typical branching fraction values calculated within the framework of perturbative
QCD~\cite{pqcd}, QCD factorization~\cite{qcdf}, soft collinear effective
theory~\cite{scet} and $SU(3)$ flavor symmetry~\cite{su3f} are in the range
$1.2$--$6.3\%$. In the past, Belle~\cite{belle-res} and $\babar$~\cite{babar-res}
have searched for $B^0\to\eta^{\prime}K^{\star}(892)^0$ with the latter reporting
the first evidence at a $4\sigma$ level. So far, $\CP$ violation has not been
probed in the decay.

We search for $B^0\to\eta^{\prime}K^{\star}(892)^0$~\cite{res5} using a data
sample of $772\times10^6$ $B\Bbar$ pairs recorded at the $\Y4S$ resonance.
The decay candidates are reconstructed from the subsequent decay modes
$\eta^{\prime}\to\eta\pi^+\pi^-$, $\eta\to\gamma\gamma$ and $K^{\star}(892)^0
\to K^+\pi^-$. Based on an extended maximum likelihood fit to their distributions
of $M_{\rm bc}$, $\Delta E$, continuum suppression variable, and the cosine of
the $K^{\star}$ helicity angle, we extract a signal yield of $31\pm9$ events
with a significance of $5\sigma$, including systematic uncertainties. This
constitutes the first observation of the decay channel. The yield is translated
to a branching fraction ${\cal B}[B^0\to\eta^{\prime}K^{\star}(892)^0]=[2.6\pm
0.7\stat\pm0.2\syst]\times10^{-6}$, in a good agreement with theory predictions.
We also measure direct $\CP$ violation by splitting the obtained yield according
to the flavor of the $B$ meson, based on the sign of the daughter kaon from the
$K^{\star}$ decay. The obtained result, $A_{\CP}=-0.22\pm0.29\stat0.07\syst$,
is consistent with no $\CP$ violation.

\section{\boldmath An amplitude analysis of $B\to\jpsi K\pi$}
\label{Sec-7}

Recently, a number of new charmonium-like states have been observed at
the $B$ factories and elsewhere. Some of them especially the charged ones
look very much like exotic, defying predictions of the quark model. The
first one in the series, the $Z_c(4430)^+$, was discovered by Belle in the
$\psitwos\pi^+$ invariant mass spectrum in $B^0\to\psitwos K^-\pi^+$~\cite{zc1},
followed by two more states, the $Z_c(4050)^+$ and $Z_c(4250)^+$, in $B^0\to
\chi_{c1}K^-\pi^+$ decays~\cite{zc2}. Lately, BESIII has joined the game by
observing $Z_c(3900)^+$ in the $\jpsi\pi^+$ invariant mass spectrum in $Y(4260)
\to\jpsi\pi^+\pi^-$~\cite{zc3}. In a back-to-back publication~\cite{zc4},
Belle corroborated the finding.

Motivated by these exciting results, we perform an amplitude analysis~\cite{res6}
of the decay $B^0\to\jpsi K^-\pi^+$, with $\jpsi\to\mu^+\mu^-$ or $\ep\en$, using
a data sample of $772\times10^6$ $B\Bbar$ pairs. Our analysis strategy is similar
to a recent study of $B^0\to\psitwos K^-\pi^+$~\cite{zc5}. In addition to the known
$Z_c(4430)^+$ state, we find a new charmonium-like state $Z_c(4200)^+$ in the
$\jpsi\pi^+$ invariant mass spectrum with a significance exceeding $6\sigma$.
The minimal quark content of this state is exotic: $|c\bar{c}u\bar{d}\!\!>$. Its
mass and decay width are measured to be $[4196^{+31}_{-29}\stat^{+17}_{-13}\syst]
\mevcc$ and $[370\pm70\stat^{+70}_{-132}\syst]\mev$, respectively. The preferred
spin-parity quantum numbers are $J^P=1^+$.

\section{Summary and outlook}
\label{Sec-8}

Though close to five years have passed by since its data taking, Belle continues
to produce high quality results. A small sample of those based on the full
statistics are presented here. That includes: a) first observation of
$D^0$-$\Dbar^0$ mixing using $D\to K\pi$ decays in $\ep\en$ collisions, b) a $2.5
\sigma$ evidence for charm mixing and no hint for $\CP$ violation in $D^0\to\KS
\pi^+\pi^-$, c) an order-of-magnitude improvement over the previous results for
$A_{\CP}$ in the decay $D^0\to\pi^0\pi^0$, d) a $1.8\sigma$ difference with respect
to the SM prediction in the lepton forward-backward asymmetry at low $q^2$ in
the inclusive $B\to X_S\ell^+\ell^-$ decays, e) first observation of the charmless
hadronic decay $B^0\to\eta^\prime K^\star(892)^0$, and f) observation of a new
charged charmonium-like state in $B^0\to\jpsi K^-\pi^+$. Such kind of unique
explorations at the next-generation $\ep\en$ flavor factory will continue with
the upcoming Belle II experiment~\cite{phill}.





\end{document}